\documentclass[a4paper,11pt]{article}
\pdfoutput=1 

\usepackage{jcappub} 
\usepackage{subcaption}
\usepackage{graphicx}
\usepackage{caption}
\usepackage{afterpage}
\usepackage{makecell}
\usepackage[normalem]{ulem}
\usepackage{multirow}
\usepackage{lscape}
\usepackage{cprotect}
\usepackage{bbm}
\usepackage{fancyvrb}
\usepackage{bm}

\title{\boldmath Radio Galaxy Zoo: Leveraging latent space representations from variational autoencoder}

\author[a,b,1]{Sambatra Andrianomena,\note{Corresponding author.}}
\author[c]{Hongming Tang}

\affiliation[a]{South African Radio Astronomy Observatory (SARAO), Black River Park, Observatory, Cape Town, 7925, South Africa}
\affiliation[b]{Department of Physics \& Astronomy, University of the Western Cape, Bellville, Cape Town 7535,
South Africa}
\affiliation[c]{Department of Astronomy, Tsinghua University, Beijing, 100084, China}

\emailAdd{andrianomena@gmail.com}

\abstract{We propose to learn latent space representations of radio galaxies, and train a very deep variational autoencoder (\protect\Verb+VDVAE+) on RGZ DR1, an unlabeled dataset, to this end. We show that the encoded features can be leveraged for downstream tasks such as classifying galaxies in labeled datasets, and similarity search. Results show that the model is able to reconstruct its given inputs, capturing the salient features of the latter. We  use the latent codes of galaxy images, from MiraBest Confident and FR-DEEP NVSS datasets, to train various non-neural network classifiers. It is found that the latter can differentiate FRI from FRII galaxies achieving \textit{accuracy} $\ge 76\%$, \textit{roc-auc} $\ge 0.86$, \textit{specificity} $\ge 0.73$ and \textit{recall} $\ge 0.78$ on MiraBest Confident dataset, comparable to results obtained in previous studies. The performance of simple classifiers trained on FR-DEEP NVSS data representations is on par with that of a deep learning classifier (CNN based) trained on images in previous work, highlighting how powerful the compressed information is. 
We successfully exploit the learned representations to search for galaxies in a dataset that are semantically similar to a query image belonging to a different dataset. Although generating new galaxy images (e.g. for data augmentation) is not our primary objective, we find that the \protect\Verb+VDVAE+ model is a relatively good emulator.  
Finally, as a step toward detecting anomaly/novelty, a density estimator -- Masked Autoregressive Flow (\protect\Verb+MAF+) -- is trained on the latent codes, such that the log-likelihood of data can be estimated. The downstream tasks conducted in this work demonstrate the meaningfulness of the latent codes.
}

\begin{document}
\maketitle
\flushbottom

\section{Introduction}
\label{introduction}
Galaxy morphology is a powerful probe for investigating galaxy evolutionary processes, e.g. star formation history, the physical processes that galaxies undergo in their environment. Surveys like DESI \cite{dey2019overview} and SDSS \cite{gunn1998sloan, gunn20062}, which make tens of millions of galaxy images available, provide insights into galaxy formation and evolution. On the radio counterpart, a great deal of effort has been made toward building datasets of radio galaxy images, e.g. Radio Galaxy Zoo \cite{banfield2015radio}, and upcoming large experiments like SKA \cite{dewdney2009square, weltman2020fundamental} will increase the amount of data available. Most of the methods that have been considered to identify galaxies with different morphological features are supervised learning based, which heavily relies on labeling of the data. So far, they have been successful, although manual labeling process is not only expensive but could also potentially introduce biases in the data. Morever, for new scientific discoveries and searching for anomalies in large uncurated datasets, resorting to the feature extractors that are trained in a supervised learning setup is not optimal due to the fact that they are not robust to both noise and dataset shift. 

Self-supervised learning (\verb|SSL|) \cite{chen2020simple, grill2020bootstrap, he2020momentum, chen2021exploring}, which does not require data labeling, has been considered to uncover patterns in unlabeled dataset by learning robust representations of the high dimensional images. For example, \cite{stein2021self} successfully used constrastive learning to search for galaxies that are semantically similar in large datasets. \cite{hayat2021self} considered \verb|SimCLR| method \cite{chen2020big} to learn representations of astronomical images from SDSS, and \cite{slijepcevic2022learning} opted for Boostrap Your Own Latent (\verb|BYOL|) method \cite{grill2020bootstrap} to extract important features of radio galaxies. 

In this work, we aim to learn latent codes of radio galaxies using a generative model, Very Deep Variational AutoEncoder (\verb|VDVAE|). Earlier work \cite{bastien2021structured} used \verb|VAE|, whose both encoder and decoder were composed of only fully connected layers, to generate synthetic images of Fanaroff-Riley Class I (FRI) and Class II (FRII) radio galaxies. Their approach was capable of generating realistic radio galaxy images, but the generated and reconstructed images were blurry, which could be attributed to the lack of expressivity of the network. Our main goal in this work, unlike the case studied in \cite{bastien2021structured}, is to highlight the ability of a deep generative model, \verb|VDVAE|, to learn meaningful representations which can be leveraged for various downstream tasks. We also show how to estimate the log-likelihood of data using the learned representations, which is useful within the context of anomaly/novelty detection. We present the datasets used in our analyses in Section \ref{sec:data}, and introduce the model considered in this study and other \verb|SSL| based methods used for comparison in Section \ref{algo}. The main results and the data likelihood estimation are reported in Sections \ref{sec:results} and \ref{sec:likelihood} respectively, and we conclude in Section \ref{sec:conclusion}.

\section{Data}\label{sec:data}
We make use of the Radio Galaxy Zoo Data Release 1 (RGZ DR1) (Wong et al. 2023 in prep)
to train and evaluate our generative model. The dataset used in our analyses contains $\sim 100,000$ unlabeled galaxies with their corresponding projected angular size in arcseconds. The input image to our model has a selected resolution of $64\times64$ pixels. \\
 To investigate the ability of our network, and that of other \verb|SSL| based methods used for comparison in our analyses, to compress the images, we train various non-neural network methods on the latent features of galaxy images from two different datasets, MiraBest \textit{Confident} dataset (MBC)\footnote{The data can be obtained from \href{https://github.com/as595/E2CNNRadGal/tree/main}{https://github.com/as595/E2CNNRadGal/tree/main}.}\cite{miraghaei2017nuclear, porter2020mirabest, porter2023mirabest} and FR-DEEP NVSS dataset \cite{tang2019transfer}\footnote{The data can be downloaded from \href{https://github.com/HongmingTang060313/FR-DEEP}{https://github.com/HongmingTang060313/FR-DEEP}.}. The idea is to identify FRI and FRII galaxy images in each dataset by only exploiting their representations. MBC and FR-DEEP NVSS have 729/104 (train/test) and 550/50 (train/test) instances respectively, and their images are also cropped to $64\times 64$ pixels. The numbers of FRI and FRII in the training examples are roughly equal in both datasets, with an imbalance ratio $\sim 0.5$. It is worth noting that the RGZ DR1 contains some MBC samples which are flagged out when training the feature extractors.

\section{Models}\label{algo}
In our investigation, we also train various \verb|SSL| based methods and compare their performance with that of our network, specifically in terms of using the encoded features to identify galaxy types in labeled datasets, MBC and FR-DEEP NVSS. In this section we provide the technical details of each algorithm together with the hyperparameters selected to train them. 
\subsection{Very deep variational autoencoder (\protect\Verb+VDVAE+)}
Variational autoencoder (\verb|VAE|) \cite{kingma2013auto} is a type of generative model that is composed of an encoder $q_{\phi}(\bm{z}|\bm{x})$ -- which is an approximate posterior given the intractability of the true posterior --, a decoder $p_{\theta} (\bm{x}|\bm{z})$ and a prior $p_{\theta}(\bm{z})$. The two networks $\phi$ and $\theta$ are simultaneously trained by maximizing the evidence lower bound ($\mathbb{ELBO}$) 

\begin{equation}
    \mathbb{ELBO} = E_{\bm{z}\sim q_{\phi}(\bm{z}|\bm{x})}{\rm log}p_{\theta}(\bm{x}|\bm{z}) - D_{KL}(q_{\phi}(\bm{z}|\bm{x})|| p_{\theta}(\bm{z})),
\end{equation}\label{eq:elbovae}where the first term denotes the reconstruction error which measures how well the model recovers the inputs, and the second term is the Kullback-Leibler (KL) divergence, quantifying the dissimilarity between $q_{\phi}(\bm{z}|\bm{x})$ and $p_{\theta}(\bm{z})$. 
\begin{figure}[ht]
\vskip 0.2in
\begin{center}
\centerline{\includegraphics[width=\columnwidth]{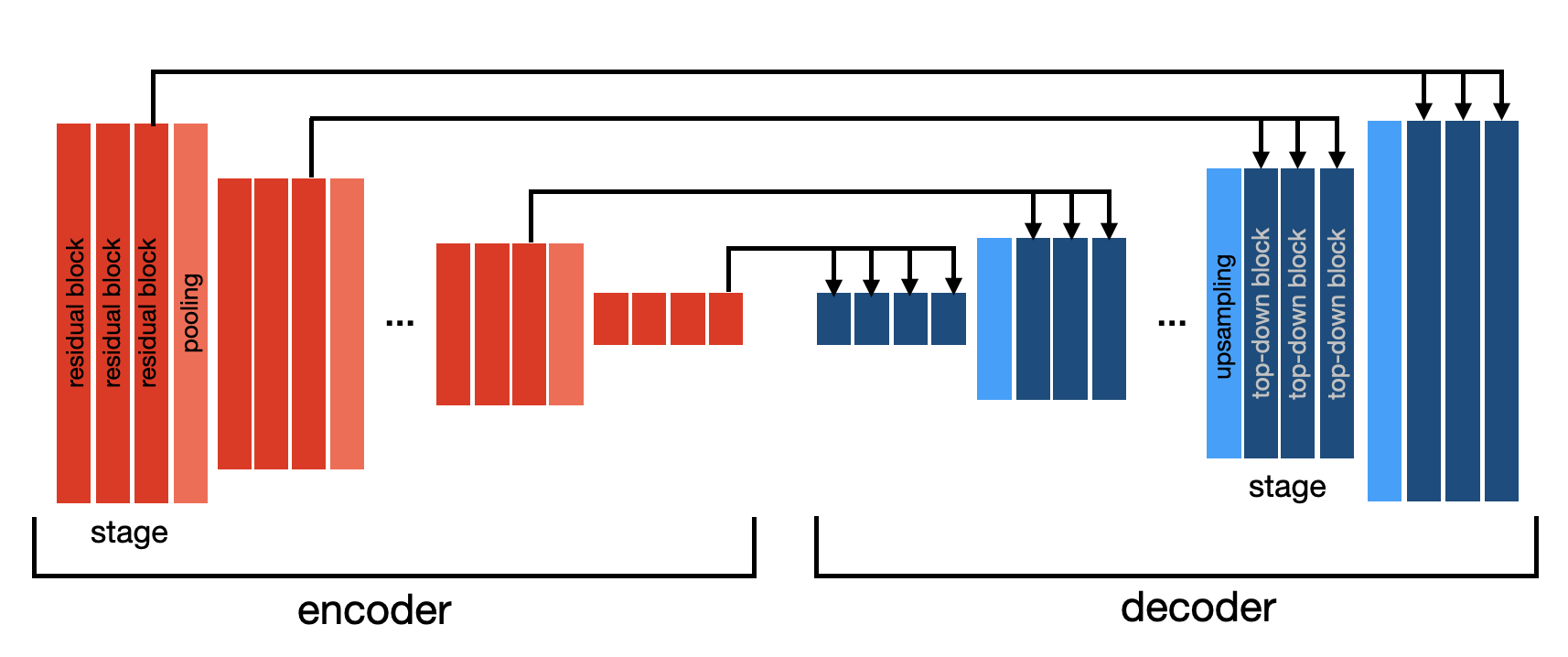}}
\cprotect\caption{Schematic diagram of the \verb|VDVAE| model. The red and blue blocks denote the residual blocks of the encoder and top-down blocks of the decoder respectively. The black arrows indicate mixing via concatenation along the channel dimension.}
\label{fig:schematic}
\end{center}
\vskip -0.2in
\end{figure}
It is worth noting that \verb|VAE| outputs (either reconstructed or generated images) are known to suffer from blurriness, which can be potentially mitigated by controlling the contribution of the KL divergence to the total loss, using a hyperparameter $\beta$ according to \cite{burgess2018understanding}
\begin{equation}
    \mathbb{ELBO} = E_{\bm{z}\sim q_{\phi}(\bm{z}|\bm{x})}{\rm log}p_{\theta}(\bm{x}|\bm{z}) - \beta D_{KL}(q_{\phi}(\bm{z}|\bm{x})|| p_{\theta}(\bm{z})).
\end{equation} There are several variants of the \verb|VAE| models but we consider the Very Deep Variational Autoencoder (\verb|VDVAE|) model prescribed by \cite{child2020very} in our analyses. In order to increase the expressivity of both the prior $p_{\theta}(\bm{z})$ and approximate posterior $q_{\phi}(\bm{z}|\bm{x})$, \cite{child2020very} proposed a hierarchical \verb|VAE| comprising many stochastic layers of latent variables. The latter have different resolutions $\bm{z}_{0},\bm{z}_{1},...,\bm{z}_{N}$  which are conditionally dependent on each other according to 
\begin{eqnarray}\label{eq:conditionals}
    p_{\theta}(\bm{z}) = p_{\theta}(\bm{z}_{0})\prod_{k = 1}^{N}p_{\theta}(\bm{z}_{k}|\bm{z}_{k-1}), \nonumber\\
    q_{\phi}(\bm{z}|\bm{x}) = q_{\phi}(\bm{z}_{0}|\bm{x})\prod_{k = 1}^{N}q_{\phi}(\bm{z}_{k}|\bm{z}_{k-1}, \bm{x}),
\end{eqnarray}where $N$ is the number of layers, and the conditionals $q_{\phi}(\cdot)$ and $p_{\theta}(\cdot)$ are parameterized as diagonal Gaussians. In this work, we consider the latent variable with the lowest resolution $\bm{z}_{0}$ which is a vector of length 256, i.e. a feature vector with 256 components. Figure~\ref{fig:schematic} presents a schematic diagram of the model architecture. A residual block, which comprises 4 convolutional layers, is an important component of the two networks $\phi$ and $\theta$. The encoder contains multiple stages which are built by stacking residual blocks (see red blocks in Figure \ref{fig:schematic}). The output of one stage is downsampled by using average pooling. Each stage of the decoder is composed of chained top-down blocks. At the level of each top-down block, the prior, the posterior and the latent variable are computed by using one residual block, another residual block and one convolutional layer respectively; and a third residual block is used at the output. The feature maps outputted by the last top-down block at a given stage is upsampled using nearest neighbor method. It is noted that both networks ($\phi$ and $\theta$) have the same number of stages and the dimensions of feature maps from two corresponding stages are the same. The input of each top-down block of the decoder at a given stage is concatenated with the output of the last residual block at the corresponding stage of the encoder (see Figure \ref{fig:schematic}). The augmented feature maps resulting from this mixing are used to compute the conditionals $q_{\phi}(\cdot)$ and $p_{\theta}(\cdot)$ in Equation \ref{eq:conditionals}. The encoder and decoder have 6 stages of $\{3, 3, 2, 2, 2, 1\}$ residual blocks and $\{5, 5, 4, 3, 2, 1\}$ top-down blocks respectively. The decoder is chosen to be a bit  deeper for good quality of generated images. Our choice might be suboptimal but good enough for our purpose. We consider \verb|RMSProp| optimizer with a learning rate of 0.00002, momentum set to 0.9, and weight decay of 0.0001. We train the model for 100 epochs with batch size of 32. The learning rate is reduced by a factor 0.5 whenever the validation loss does not improve over 10 epochs during training, i.e. using \verb|ReduceLROnPlateau| scheduler. We mainly follow the parameters in \cite{child2020very} with some adjustments due to computing resources. 
\subsection{\protect\Verb+SimCLR+ method}
Contrastive learning method consists of minimizing the distance between two different augmentations of an image in latent space while increasing distance between representations of augmented views of different images, i.e. in latent space, an image and its transformations are clustered, and pushed away from other images and their corresponding augmentations.
\begin{figure}[ht]
\vskip 0.2in
\begin{center}
\centerline{\includegraphics[width=0.7\columnwidth]{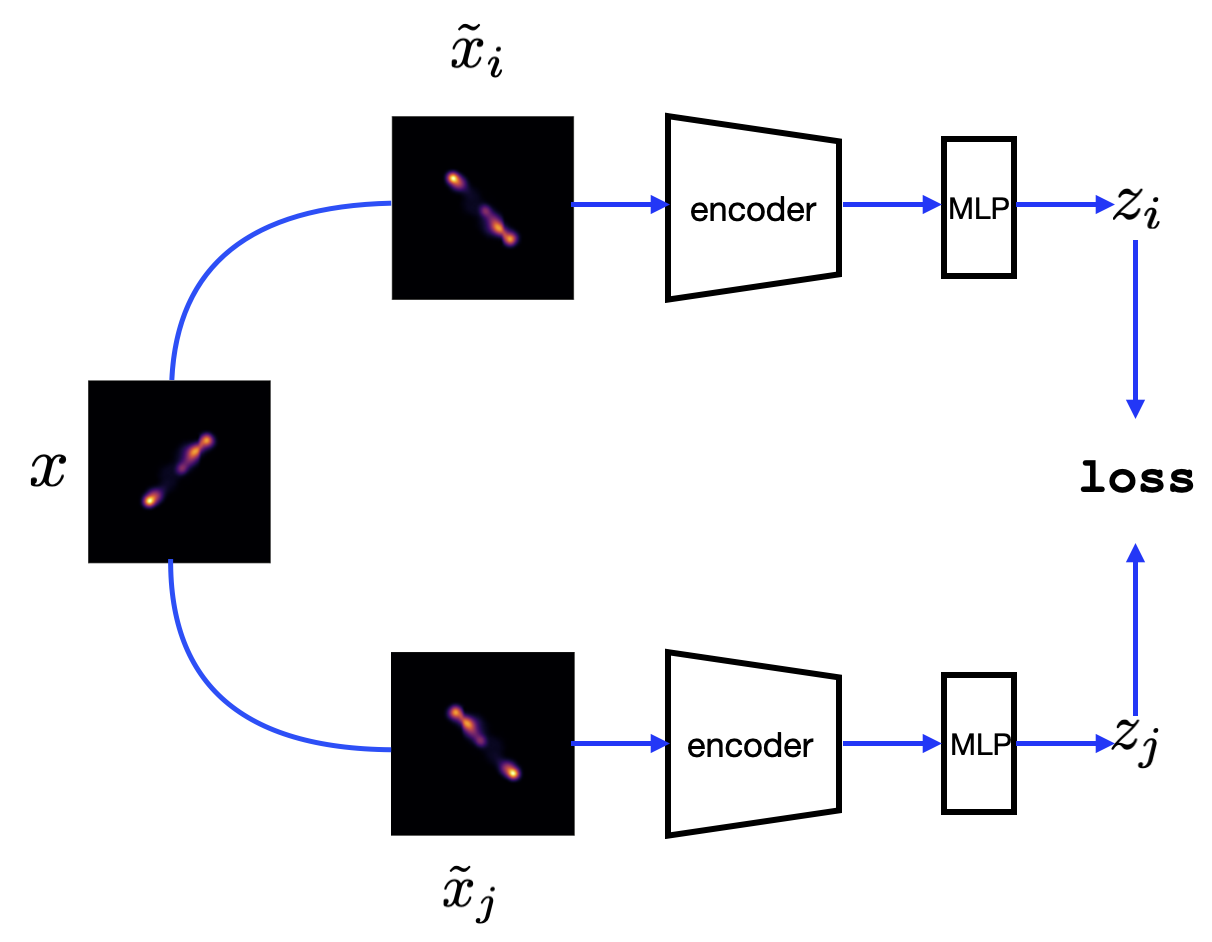}}
\cprotect\caption{Schematic diagram of \verb|SimCLR| method.}
\label{fig:simclr}
\end{center}
\vskip -0.2in
\end{figure}
In this work we consider \verb|SimCLR| method \cite{chen2020simple} which applies two stochastic transformations to an image, resulting in two different augmented views $\Tilde{\bm{x}}_{i}$ and $\Tilde{\bm{x}}_{j}$ which form a positive pair $\{\Tilde{\bm{x}}_{i}, \Tilde{\bm{x}}_{j}\}$. The corresponding features of the latter -- $\bm{h}_{i}$ and $\bm{h}_{j}$ respectively -- are extracted via an encoder. Finally, the representations $\{\bm{h}_{i}$, $\bm{h}_{j}\}$ are projected into latent space using a multilayer perceptron (\verb|MLP|), giving $\{\bm{z}_{i}, \bm{z}_{j}\}$ as shown in Figure \ref{fig:simclr}. The separation of positive pairs $\{\bm{z}_{i}, \bm{z}_{j}\}$ in latent space is minimized while that of negative pairs is maximized using a constrastive loss, also known as \textit{NT-Xent} (normalized temperature-scaled cross entropy loss) \cite{oord2018representation, chen2020simple}
\begin{equation}\label{eq:infonce}
    \ell(i,j) = -{\rm log}\frac{{\rm exp}({\rm sim}(\bm{z}_{i},\bm{z}_{j}) / \tau)}{\sum_{k=1}^{2N}\mathbbm{1}_{[k \ne i]}{\rm exp}({\rm sim}(\bm{z}_{i},\bm{z}_{k}) / \tau)},
\end{equation}
where $\mathbbm{1}_{[k \ne i]}$ is equal to 1, 0 if 
and only if $k \ne i$ (for negative pairs) and $k = i$ respectively, and $\tau$ is known as the temperature parameter. The function ${\rm sim}(\bm{z}_i,\bm{z}_j)$ denotes cosine similarity ${\rm sim}(\bm{z}_i,\bm{z}_j) = \bm{z}_{i}\cdot\bm{z}_{j} / (||\bm{z}_{i}||\>||\bm{z}_{j}||)$. In our case, the stochastic transformation is defined by a set of data augmentations which are a random horizontal flip, a random vertical flip, a random crop, a random color jitter, and a Gaussian blur. We make use of \verb|Resnet-34| \cite{he2016deep} as a backbone. The model is trained for 1000 epochs, using LARS optimizer with a learning rate \verb|lr| = 0.001 and a batch of 1024 instances. The  encoded features\footnote{The outputs of the encoder which is \protect\Verb+Resnet-34+.} which are arrays of length 512 are projected into latent space using an \verb|MLP| with one hidden layer, yielding vectors with 128 components.   
\subsection{\protect\Verb+BYOL+ method}
In order to avoid collapsed representations, constrastive methods such as \verb|SimCLR| learn to distinguish representations of distorted views of an image from those of different images, and  
\begin{figure}[ht]
\vskip 0.2in
\begin{center}
\centerline{\includegraphics[width=0.8\columnwidth]{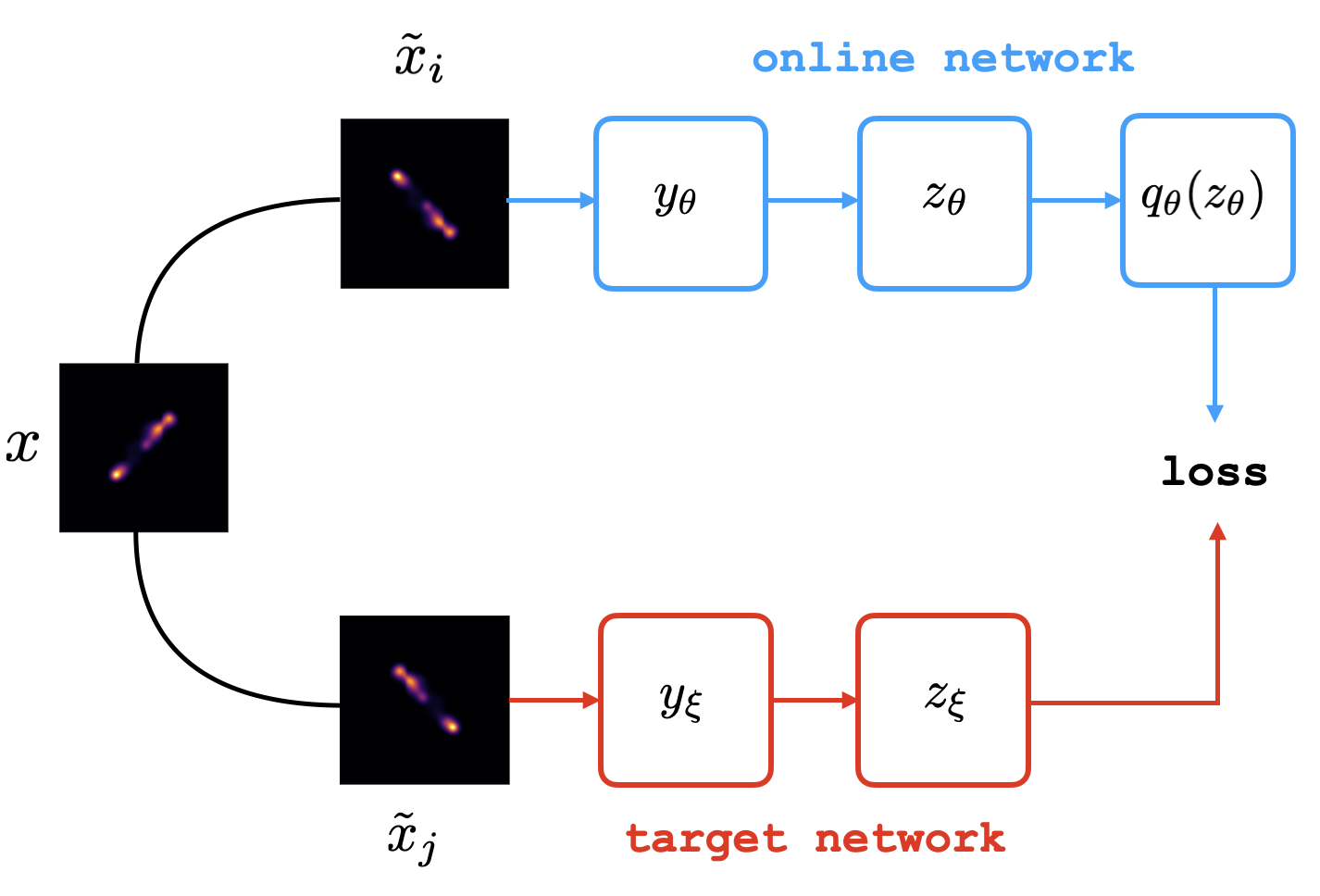}}
\cprotect\caption{Schematic diagram of \verb|BYOL| method. The \textit{online} network components are shown in blue, whereas those of the target are in red.}
\label{fig:byol}
\end{center}
\vskip -0.2in
\end{figure}
the representations learned by \verb|SimCLR| are of better quality with larger batches during training \cite{chen2020simple}. Unlike \verb|SimCLR|, \verb|BYOL| method \cite{grill2020bootstrap} bypasses the need for negative examples, but rather uses an \textit{online} network that learns to predict the outputs of a \textit{target} network (Figure~\ref{fig:byol}). The former, defined by its parameters $\theta$, comprises an encoder that outputs a representation $y_{\theta}$ which, similar to the case of \verb|SimCLR|, is projected into latent space $z_{\theta}$. To avoid collapsing results, a predictor $q_{\theta}(z_{\theta})$, which processes $z_{\theta}$, is added to the \textit{online} network (see Figure~\ref{fig:byol}). The \textit{target} network architecture is a copy of that of the \textit{online} but its weights $\xi$ are computed from an exponential average of $\theta$ at each training step according to 
\begin{equation}
    \xi \longleftarrow \kappa\xi + (1 - \kappa)\theta, 
\end{equation}
where $\kappa$ indicates the decay rate $\in$ [0, 1]. In other words, the gradients related to target parameters are not computed. Two augmented views, obtained from stochastic transformations, of an image $\Tilde{\bm{x}_{i}}$ and $\Tilde{\bm{x}_{j}}$ are passed through the online and target pipelines (see Figure~\ref{fig:byol}) respectively, and the \textit{online} network is trained to predict the target $z_{\xi}$, resulting in refined representations. This boostrapping procedure helps the \textit{online} network improve the quality of its learned representations as the training progresses. The loss is defined by a mean squared error between the target projection and the online prediction \cite{grill2020bootstrap}
\begin{equation}
    \mathcal{L} = ||q_{\theta}(z_{\theta}) - z_{\xi}||^{2}_{2}.
\end{equation}
We also consider \verb|Resnet-34| as backbone, and opt for LARS optimizer with learning rate of $0.0005$, batch of 1024 examples and training epochs of 1000. Both online projector and predictor consist of one hidden layer \verb|MLP|. The set of data augmentations considered during training is the same as the one for \verb|SimCLR|.
\subsection{\protect\Verb+SimSiam+ method}
\begin{figure}[ht]
\vskip 0.2in
\begin{center}
\centerline{\includegraphics[width=1\columnwidth]{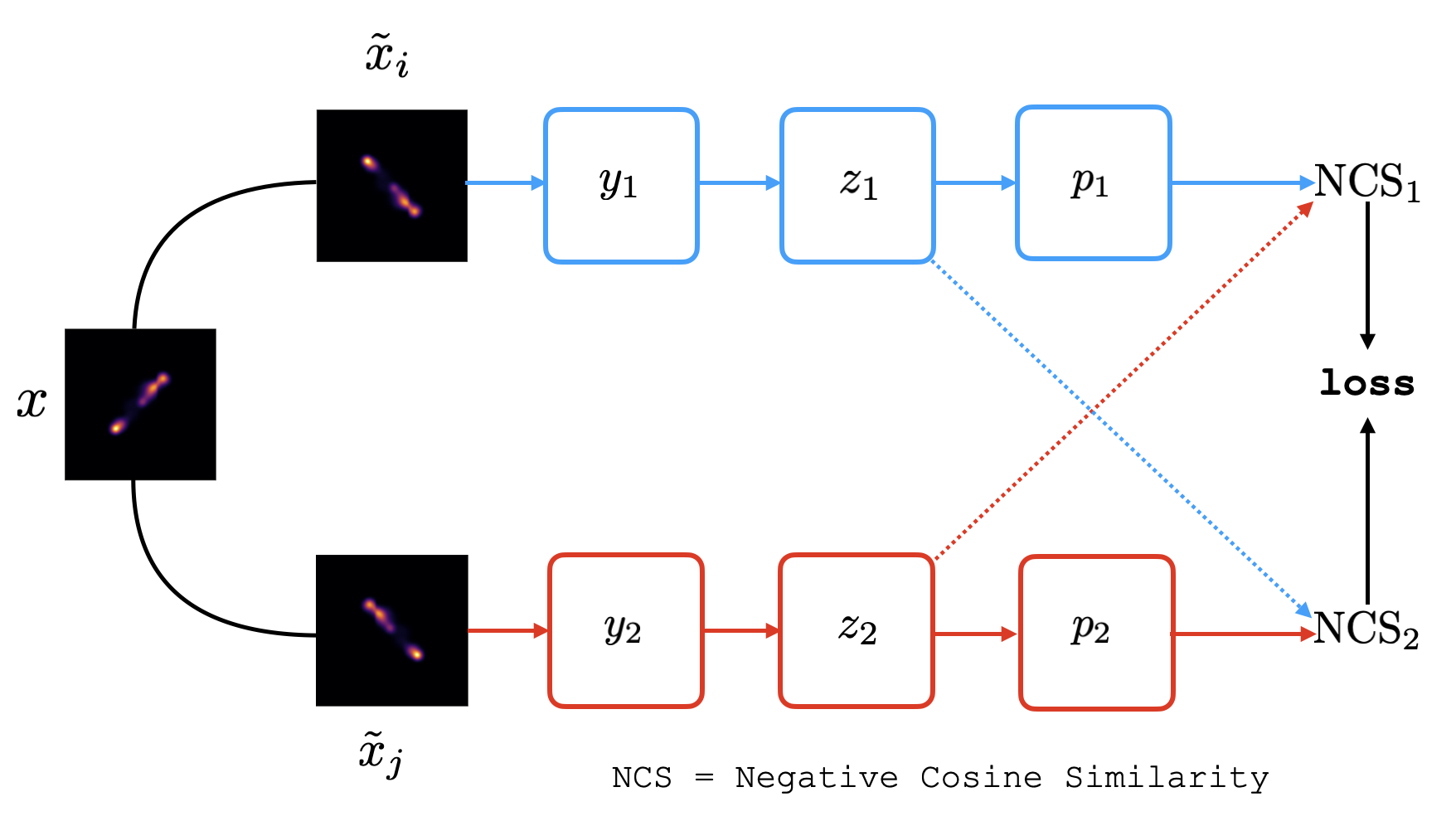}}
\cprotect\caption{Schematic diagram of \verb|SimSiam| method. In the first blue branch, the prediction $p_1$ is matched with the representation $z_{2}$ using negative cosine similarity, indicated by ${\rm NCS}_{k = 1,2}$. The dashed red arrow indicates that $z_{2}$ acts like a constant with zero gradient. Similarly in the second red branch, $p_2$ is matched with $z_1$ which is turned into a constant, as indicated by the dashed blue arrow.}
\label{fig:simsiam}
\end{center}
\vskip -0.2in
\end{figure}

\verb|SimSiam| \cite{chen2021exploring} rejects the need for a momentum encoder and negative examples altogether to prevent collapsing results. Like \verb|SimCLR|, the parameters are shared between the two pipelines (blue and red branches in Figure~\ref{fig:simsiam}), 
and similar to \verb|BYOL|, an augmented view of an image is predicted from another augmented view of the same image. In \verb|SimSiam|, two different augmentations of the same image $\Tilde{\bm{x}}_{i}$ and $\Tilde{\bm{x}}_{j}$ are encoded to obtain two representations $y_{1}$ and $y_{2}$ respectively. The latter are in turn projected into a latent space, producing $z_{1}$ and $z_{2}$ respectively. The prediction $p_{1}$, which results from transforming $z_{1}$ via a projection head, is matched with the latent space representation $z_{2}$ of the second branch, by minimizing the negative cosine similarity \cite{chen2021exploring} 
\begin{equation}\label{ncs}
    \mathcal{D}(p_1, z_2) = - \frac{p_1}{||p_1||_{2}}\cdot \frac{z_2^{\rm stopgrad}}{||z_2^{\rm stopgrad}||_{2}},
\end{equation}
where $z_2^{\rm stopgrad}$ denotes stop-gradient operation on $z_{2}$, which is the key aspect of the method. The prediction $p_{2}$ from the second branch is similarly matched with $z_{1}$ on which stop-gradient is acting as well and the total loss is given by \cite{chen2021exploring}
\begin{equation}
    \mathcal{L} = \frac{1}{2}(\mathcal{D}(p_1, z_2) + \mathcal{D}(p_2, z_1)).
\end{equation}
We select \verb|Resnet-34| as the encoder in our implementation, and use LARS optimizer with learning rate 0.0005. We train the model for 1000 epochs, and choose a batch size 1024. Both the projector and predictor are one hidden layer \verb|MLP| that converts their input into an array of length 128. We also use the same stochastic transformations chosen in the case of both \verb|SimCLR| and \verb|BYOL|. It is worth noting that we use \verb|lightly ssl| \cite{susmelj2020lightly} framework and follow the examples in their documentation\footnote{\href{https://docs.lightly.ai/self-supervised-learning/}{https://docs.lightly.ai/self-supervised-learning/}} to build the architecture of all the \verb|SSL| based models in this work.
\section{Results}\label{sec:results}
\begin{figure}[ht]
\vskip 0.2in
\begin{center}
\centerline{\includegraphics[width=\columnwidth]{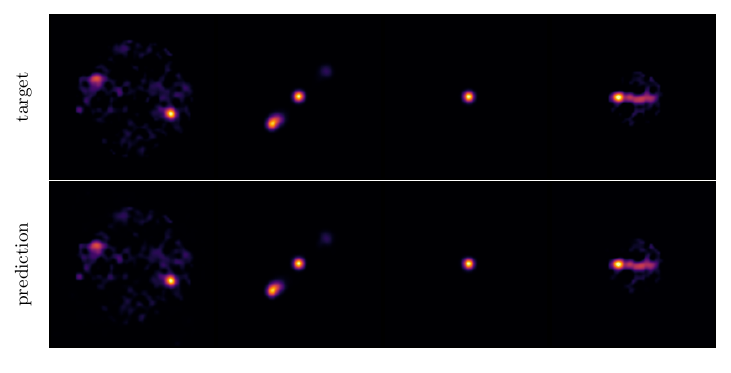}}
\caption{Reconstructing some examples of images from the test set from RGZ DR1 dataset. Top row denotes the images from the test set and the bottom row shows the corresponding images (i.e. output image of the decoder when feeding an input image) that are recovered by the decoder.}
\label{reconstruction}
\end{center}
\vskip -0.2in
\end{figure}
\subsection{Reconstruction}
\begin{figure}[ht]
\vskip 0.2in
\begin{center}
\centerline{\includegraphics[width=1.05\columnwidth]{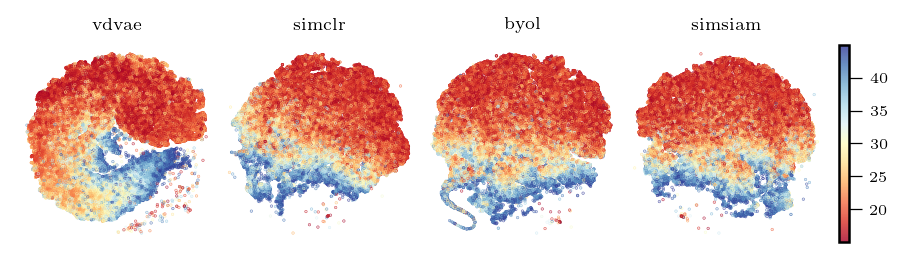}}
\cprotect\caption{Latent space representations of galaxies from the training set learned by different methods, \verb|VDVAE|, \verb|SimCLR|, \verb|BYOL|, and \verb|SimSiam|. For visualisation, TSNE method is used.  The color coding indicates the angular scale of the galaxy in arcseconds
.}
\label{features-2d}
\end{center}
\vskip -0.2in
\end{figure}
The top and bottom rows in Figure \ref{reconstruction} show some examples of input images from the unlabeled test set of RGZ DR1 dataset and their corresponding reconstructions by the decoder respectively. Results suggest that the model is able to reconstruct the targets. \verb|VAE| is known to suffer from blurry generated/reconstructed images, and the examples presented in Figure \ref{reconstruction} have been cherry-picked to highlight the predictive power of the algorithm which can recover a diffuse jet of a target (e.g. last right panel of the bottom row). 
It can be noticed that visually the diffuse structures surrounding the hot spots in the first left panel top row are captured (see first left panel of the bottom row). 
Overall, all the fine details of the inputs are reconstructed reasonably well.  
\subsection{Latent codes}\label{latents-code}
\subsubsection{Visualization of the learned representations}
\begin{figure}[ht]
\vskip 0.2in
\begin{center}
\centerline{\includegraphics[width=0.7\columnwidth]{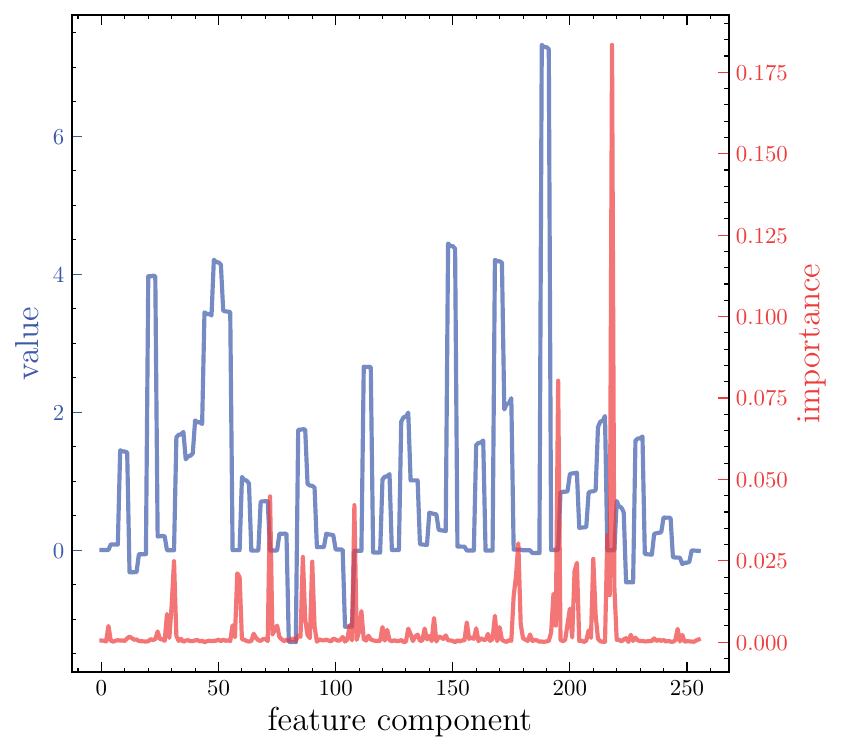}}
\cprotect\caption{The blue and red lines denote the value and feature importance score as a function of feature component respectively. The representations are those learned by \verb|VDVAE|.}
\label{fig:feature-importance}
\end{center}
\vskip -0.2in
\end{figure}
The entire training dataset is fed to each encoder in order to extract the representations which consist of vector of length 256 for the case of \verb|VDVAE| and 512 for the \verb|SSL| methods since they all use similar backbone, i.e. \verb|Resnet-34|. For visualization, dimensionality reduction method is used to further project the encoded features into two dimensional subspace. We consider t-distributed stochastic neighbor embedding
(TSNE) \cite{van2008visualizing} in our analyses to demonstrate the ability of each representation learning model to compress the galaxy images. The first, second, third and fourth panels in Figure \ref{features-2d} show the results obtained from \verb|VDVAE|, \verb|SimCLR|, \verb|BYOL|, and \verb|SimSiam| respectively. Each data point in each panel in Figure \ref{features-2d} denotes the compression of each input image. The color coding indicates the projected angular size of the galaxies. Figure \ref{features-2d} shows that in general each method has learned good representations, as evidenced by the clustering of galaxies with similar angular scales in the 2D subspace. This already points to the fact that the performance of our generative model is on par with that of the selected \verb|SSL| based methods. To analyze the features extracted by the generative model, we compute their importance. Provided that the RGZ DR1 dataset is unlabeled, but only the angular scales are given, we compute the feature importance using random forest regressor by building a mapping between the latent codes and angular scales. What we address here is whether the value of a component correlates with its importance in a specific setup, which is regression in our case, given the dataset.  As the number of instances in the training set ($\sim$ 100,000) is relatively large for the algorithm, we train a random forest regressor with an inital number of estimators on the latent codes in batch of 1000. The number of estimators is increased by one when training with the next new batch. The results are presented in Figure \ref{fig:feature-importance}. The solid blue line is the average value of each component of all the examples, whereas the solid red one denotes the importance (which is a score) of each component as outputted by the algorithm after training. Figure \ref{fig:feature-importance} shows that relatively few components carry information that is useful for inferring the angular scale of a galaxy image. In fact, by using Principal Components Analysis (\verb|PCA|), we find that only two and four components encode $95\%$ and $98\%$ of the variance respectively. Figure \ref{fig:feature-importance} clearly shows that higher  value of a feature component does not correlate with its importance for this regression task.

\subsubsection{Using encoded features to classify galaxies}
The trained encoders are used to extract the features of the galaxy images from both MBC and FR-DEEP NVSS, two labeled datasets that haven't been seen by the models during training. Leveraging the latent codes, FRI and FRII galaxies in both datasets are classified by using 
a variety of non-neural network algorithms -- $k$-nearest neighbors (\verb|knn|), random forest (\verb|rf|), support vector machine (\verb|svm|), logistic regression (\verb|lr|), gradient boosting (\verb|gb|) and extra trees (\verb|ext|). We use \verb|scikit-learn| \cite{pedregosa2011scikit} to implement the classifiers whose hyperparameters are presented in Table~\ref{tab:hyperparams}

\begin{table}[h!]
 \centering
 \begin{tabular}{|c|c|}
  \hline
   method & hyperparameters \\
  \hline
  \verb|knn| & number of neighbours: 20 \\[2pt]
  \verb|rf| & number of base estimators: 260\\[2pt]
  \verb|svc| & kernel: rbf; $\gamma$: 0.2; C: 100\\[2pt]
  \verb|lr| & maximum iteration: 1000\\[2pt]
  \verb|grad| & number of base estimators: 250 \\[2pt]
  \verb|ext| & number of base estimators: 400\\[2pt]
  \hline
 \end{tabular}
 \cprotect\caption{For each method, the presented hyperparameters are the ones that are different from their default values in \verb|scikit-learn|.}
 \label{tab:hyperparams}
\end{table}

The metrics which are used to assess the classification performance of each method considered in this work are
\begin{itemize}
    \item \textit{accuracy} which is a percentage of the number of true prediction in the test set,
    \item \textit{roc-auc}, also known as the degree of separability. In other words the ability of a classifier to differentiate between the classes.
    \item \textit{recall} (or \textit{sensitivity}), describing how well the algorithm minimizes the false negative,
    \item \textit{specificity} which is a complement of \textit{recall} and says how well the negative samples are predicted.
\end{itemize}
It is noted that when computing the metrics, FRII galaxies are the positive classes and FRI the negative ones. However, since the goal is to be able to differentiate between FRI and FRII, we aim at maximizing both \textit{recall} and \textit{specificity} which is equivalent to \textit{recall} in case where FRI is considered as positive class. For this downstream task, the representations of a training set of a dataset (e.g. MBC), which are obtained from a given feature extractor (e.g. \verb|VDVAE|) are used to train various classifiers which are then tested on the representations of a test set of the same dataset. We adopt the same procedure for testing all feature extractors on all labeled datasets. The results are shown in Table \ref{tab:lineartest}. \\
On MBC dataset, results suggest that overall the representations learned by \verb|VDVAE|, compared to those by \verb|SSL| methods, carry a bit more information such that \verb|ext| classifier generalizes  better, achieving \textit{accuracy} of $82\%$ and \textit{roc-auc} of 0.90. Moreover, both FRI and FRII are equally well classified, as evidenced by \textit{specificity} and \textit{recall} both equal to 0.82. The second, third, and fourth best classifiers, namely by \verb|rf|, \verb|grad| and \verb|knn| respectively, on \verb|VDVAE| derived representations outperform all the best classifiers (performance written in bold in Table \ref{tab:lineartest}) resulting from training on the \verb|SSL| extracted representations. This further demonstrates the better quality of the latent codes (i.e. obtained from \verb|VDVAE|). \cite{slijepcevic2022learning} and \cite{slijepcevic2023radio}, both resorting to \verb|BYOL| to learn the galaxy image representations from RGZ DR1, showed that by setting a threshold cut on the angular extent of the galaxies in RGZ DR1 (essentially removing the point source looking images from the training set) their \verb|knn| achieved better \textit{accuracy} $85.25\%$ as opposed to the case which includes all instances in RGZ DR1 when training their \verb|BYOL|. We find that the performance of our \verb|knn| on classifying the representations of MBC dataset, obtained from \verb|VDVAE|, is similar to that of \verb|knn| in \cite{slijepcevic2022learning} where a threshold cut of about 16 arcsec was adopted. And the ability of our \verb|ext| method ($82\%$ \textit{accuracy}) to classify MBC galaxies is on par with that of \verb|knn| ($85.25\%$ \textit{accuracy}) in \cite{slijepcevic2022learning} where 29 arcsec threshold was adopted.\\
On FR-DEEP NVSS dataset, it appears that the top classifiers in all setups perform equally well, with a slight advantage of \verb|lr| method classifying the representations obtained from \verb|SimCLR|. Interestingly, a simple logistic regression generalizes well on the \verb|SSL| extracted representations overall, indicating a linear mapping between the targets and the learned features. \cite{tang2019transfer} used deep CNN architecture whose weights had been previously trained on a different galaxy dataset for classification \cite{aniyan2017classifying}, an approach known as \textit{transfer learning} which can be exploited when the number of training examples of a new task is relatively small. Their deep network achieved an \textit{accuracy} of $73\%$, and \textit{roc-auc} of 0.81, \textit{specificity} $\sim 71\%$ and \textit{recall} $\sim 88\%$ on FR-DEEP NVSS data. In comparison, all our top classifiers in all setups exhibit similar performance if not better. This demonstrates how relevant and powerful the compressed information is.   
\begin{table}
  \centering
  \renewcommand{\arraystretch}{1.2}
  \begin{tabular}{|p{2cm}|c|c|c|c|c|c|c|c|}
    \hline
    \multirow{2}{2cm}{Algorithm} & \multicolumn{4}{c|}{MBC} &  \multicolumn{4}{c|}{FR-DEEP NVSS}\\
    \cline{2-9}
    & \textit{acc} & \textit{roc} & \textit{spec} & \textit{rec} & \textit{acc} & \textit{roc} & \textit{spec} & \textit{rec}\\
    \hline
    \multicolumn{9}{|c|}{\protect\Verb+VDVAE+} \\
    \hline
    \verb|knn| & 0.76 & 0.86 & 0.73 & 0.78 &0.74 &0.83& 0.55& 0.89 \\ 
    \verb|rf| & 0.80 & 0.89 & 0.80 & 0.80 & 0.80 & 0.83 & 0.73& 0.86\\
    \verb|svc|& 0.75& 0.88 & 0.69& 0.80  & 0.72 & 0.84 & 0.55& 0.86\\
    \verb|lr| & 0.73& 0.82 & 0.63& 0.82 & 0.72 & 0.85 & 0.50 & 0.89\\
    \verb|grad| & 0.80 & 0.89 & 0.76 & 0.84 &  0.76 & 0.86 & 0.64 & 0.86\\
    \verb|ext| & \textbf{0.82} & \textbf{0.90} &\textbf{0.82}& \textbf{0.82} & \textbf{0.80}& \textbf{0.87}&\textbf{0.73}& \textbf{0.86}\\
    \hline
    \multicolumn{9}{|c|}{\protect\Verb+SimCLR+} \\
    \hline
    \verb|knn| & 0.63 & 0.72 & 0.57 & 0.69 & 0.72 &0.78 & 0.45 & 0.93 \\ 
    \verb|rf| & 0.67 &0.76& 0.55 & 0.78& 0.74 & 0.82 & 0.50 & 0.93\\
    \verb|svc|& \textbf{0.72}& \textbf{0.82} &\textbf{0.71} & \textbf{0.73} & 0.76 &0.83 & 0.50 & 0.96\\
    \verb|lr| & 0.71 & 0.73 & 0.63 &0.78 & \textbf{0.86} & \textbf{0.84} & \textbf{0.77}& \textbf{0.93}\\
    \verb|grad| & 0.64& 0.73& 0.69 & 0.60 & 0.78 & 0.82 & 0.64 & 0.89\\
    \verb|ext| & 0.67 & 0.75 & 0.55 & 0.78 & 0.76 & 0.82 & 0.55 & 0.93\\
    \hline
    \multicolumn{9}{|c|}{\protect\Verb+BYOL+} \\
    \hline
    \verb|knn| & 0.63 & 0.72 & 0.67 & 0.60 &0.72 & 0.78 & 0.41 & 0.96 \\ 
    \verb|rf| & 0.70 & 0.77 & 0.61 & 0.78 & 0.72 & 0.84 & 0.50 & 0.89\\
    \verb|svc|& \textbf{0.73} & \textbf{0.77} & \textbf{0.73} & \textbf{0.73}  & \textbf{0.84} & \textbf{0.88} & \textbf{0.68} & \textbf{0.96}\\
    \verb|lr| & 0.70 & 0.68 & 0.61& 0.78 & 0.80 & 0.84 & 0.64 & 0.93\\
    \verb|grad| & 0.64 & 0.68 & 0.59 & 0.69 & 0.76 &0.81 & 0.55 & 0.93\\
    \verb|ext| & 0.69 & 0.77 & 0.57 & 0.80 & 0.74 & 0.83 & 0.55 & 0.89\\
    \hline
    \multicolumn{9}{|c|}{\protect\Verb+SimSiam+} \\
    \hline
    \verb|knn| & 0.59 & 0.70 & 0.61 & 0.56 &0.74 & 0.81 & 0.45 & 0.96 \\ 
    \verb|rf| & 0.66 & 0.81 & 0.57 & 0.75 & 0.76 & 0.84 & 0.55 & 0.93\\
    \verb|svc|& 0.67& 0.77 & 0.61 & 0.73 & 0.74 & 0.86 & 0.45 & 0.96\\
    \verb|lr| & 0.66 & 0.73 & 0.69 &0.64 & \textbf{0.82} &\textbf{0.83}& \textbf{0.73} &\textbf{0.89}\\
    \verb|grad| & \textbf{0.71} & \textbf{0.76}& \textbf{0.71} & \textbf{0.71} & 0.80 & 0.85 & 0.64 &0.93\\
    \verb|ext| & 0.73 & 0.80 & 0.61 & 0.84 & 0.74 & 0.83 & 0.50 &0.93\\
    \hline
  \end{tabular}
  \cprotect\caption{\textit{Accuracy} (\textit{acc}), \textit{roc-auc} (\textit{roc}), \textit{specificity} (\textit{spec}) and \textit{recall} (\textit{rec}) values obtained from MiraBest \textit{Confident} and FR-DEEP NVSS test sets for different classifiers; k-nearest neighbors (\verb|knn|), random forest (\verb|rf|), support vector machine (\verb|svc|), logistic regression (\verb|lr|), gradient boosting (\verb|grad|), extra trees (\verb|ext|). The bold font highlights the best performance on representations learned by a method. }\label{tab:lineartest}
\end{table}

\subsection{Similarity search}
Another downstream task that exploits the latent codes is similarity search, which consists of finding images within a dataset that are semantically similar to a query image, using the vector representations. If $\bm{\theta}^{\rm query}$ is the representation of the query and $\bm{\theta}^{j}$ that of any example from the dataset within which the search is conducted, the cosine similarity is given by
\begin{equation}
    S(\bm{\theta}^{\rm query}, \bm{\theta}^{j}) = \frac{\bm{\theta}^{\rm query} \cdot \bm{\theta}^{j}}{||\bm{\theta}^{\rm query}||\> ||\bm{\theta}^{j}||}.
\end{equation}
\begin{figure}[!ht]
\vskip 0.2in
\begin{center}
\centerline{\includegraphics[width=\columnwidth]{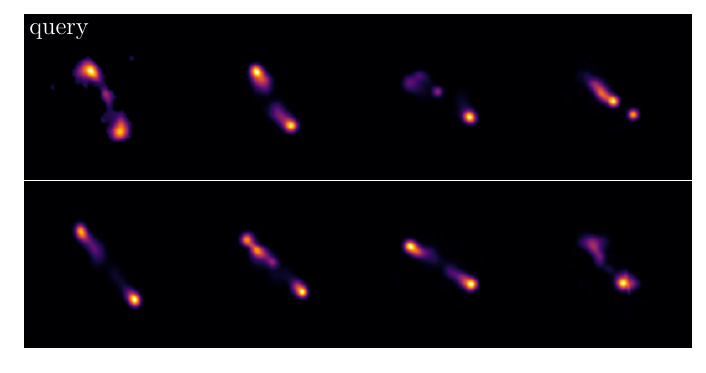}}
\caption{Similarity search exploiting the learned representations of galaxies. The top left image is the query from the test set in MBC and all the remaining images are obtained from searching in the RGZ DR1.}
\label{similarity-search}
\end{center}
\vskip -0.2in
\end{figure}
\begin{figure}[!ht]
\vskip 0.2in
\begin{center}
\centerline{\includegraphics[width=\columnwidth]{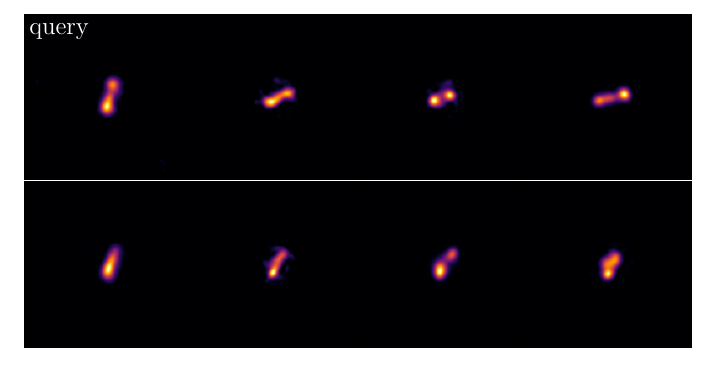}}
\caption{Similarity search where as opposed to Figure \ref{similarity-search}, the query image has a relatively small extension. The top left image is the query from the test set in MBC and all the remaining images are obtained from searching in the RGZ DR1.}
\label{similarity-search-small}
\end{center}
\vskip -0.2in
\end{figure}
The higher the score $S$ the more similar to the query an image from the dataset is. 
The query drawn from MBC dataset is used to search for galaxies which are semantically similar to it in RGZ DR1. Overall the galaxy images retrieved from the latter exhibit bright hotspots on both lobes and diffuse jets (Figure \ref{similarity-search}), which are features shared with the query shown in left panel on the top row of Figure \ref{similarity-search}.
Interestingly, all galaxies in Figure \ref{similarity-search} appear to show roughly the same inclination. The image query presented in Figure \ref{similarity-search} has larger angular extensions, so for a further test, we carry out another search for galaxies with relatively small angular extension, but bigger than a point source so that some features are visible. Similar to the previous case, the query is selected from MBC and search is conducted in RGZ DR1. Figure \ref{similarity-search-small} shows that the selected galaxies based on the query (top left panel in Figure \ref{similarity-search-small}) are semantically similar to the latter. They all roughly show diffuse emission between two bright lobes, and again are inclined in the same direction.    

\subsection{Generating new images}
\begin{figure}[ht]
\vskip 0.2in
\begin{center}
\centerline{\includegraphics[width=\columnwidth]{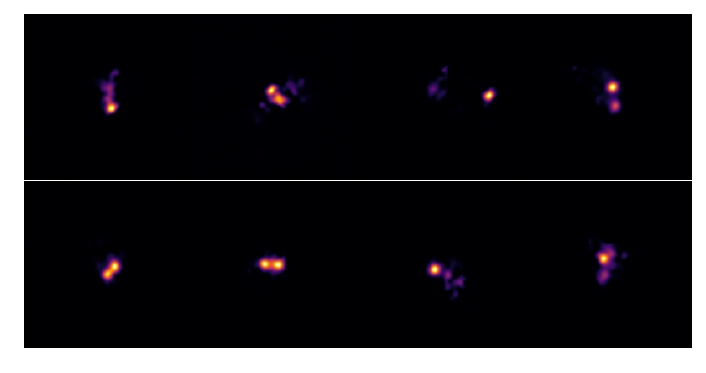}}
\caption{Examples of images that are generated by the trained decoder from sampling in the latent space.}
\label{generated}
\end{center}
\vskip -0.2in
\end{figure}

By sampling data points from the latent space and passing them through the decoder, new images are generated. We present in Figure \ref{generated} some examples of cherry-picked images that are produced by our model.
Overall, the model is able to capture the salient features of the RGZ DR1 data, such as the hotspots and diffuse structures. It can be noticed that the projected angular scales of the generated images are relatively small, similar to those of the images in Figure \ref{similarity-search-small} overall. It can be argued that this is due to the fact that the training dataset is strongly biased toward images with small angular size, as $\sim 70\%$ of the galaxies has less or equal than $35^{''}$ extension. It should be reiterated that the main objective is toward more compressing the data rather than the ability to generate new images (e.g. for data augmentation). But one possible solution, in order to reduce the effect of this bias in the generated images, is to train the generative model with a well balanced training set which contains roughly equal number of images with small and large angular scales. To further improve the quality of the generated images, the model can be conditioned on the angular extensions. We defer this to future work.

\section{Estimating log-likelihood}\label{sec:likelihood}
We have seen in Section~\ref{sec:results} that the latent codes carry meaningful information that can be exploited for some downstream tasks. The model parameters are optimized by maximizing the $\mathbb{ELBO}$ which is a lower limit of the log-likelihood. As such, estimating the log-likelihood of an input (or an entire dataset), within the context of identifying an out-of-distribution sample, is required. One way to address that is to directly train a density estimator on the $64\times64$ pixels images. However, provided the usefulness of the latent representations with smaller dimensions compared with the images, they can also be used to train a density estimator so as to estimate the log-likelihood. In this section, we opt for the latter approach and train a Masked Autoregressive Flow (\verb|MAF|) \cite{papamakarios2017masked} -- a state of the art density estimator --  on the representations. We consider \verb|denmaf| library \cite{lo2023denmarf} in our analyses, and first give a brief overview of normalizing flow and the \verb|MAF| method before presenting the results. 

\subsection{Masked Autoregressive Flow (\protect\Verb+MAF+)}
Normalizing flow \cite{rezende2015variational} is a type of generative model which consists of building an invertible differentiable mapping $f$: $\textbf{u} \rightarrow \textbf{x}$  between a data distribution $\textbf{x}\sim p(\textbf{x})$ and a base density $\textbf{u}\sim\pi_{u}(\textbf{u})$ (also known as prior) which is generally Gaussian. Using the change of variable formula, we have that \cite{papamakarios2017masked}
\begin{equation} \label{eq:change-of-variable}
p(\textbf{x}) =  \pi_{u}(f^{-1}(\textbf{x})) \left|{\rm det}\left(\frac{\partial f^{-1}}{\partial \textbf{x}}\right)\right|.
\end{equation}
\begin{figure}[ht]
\vskip 0.2in
\begin{center}
\centerline{\includegraphics[width=0.7\columnwidth]{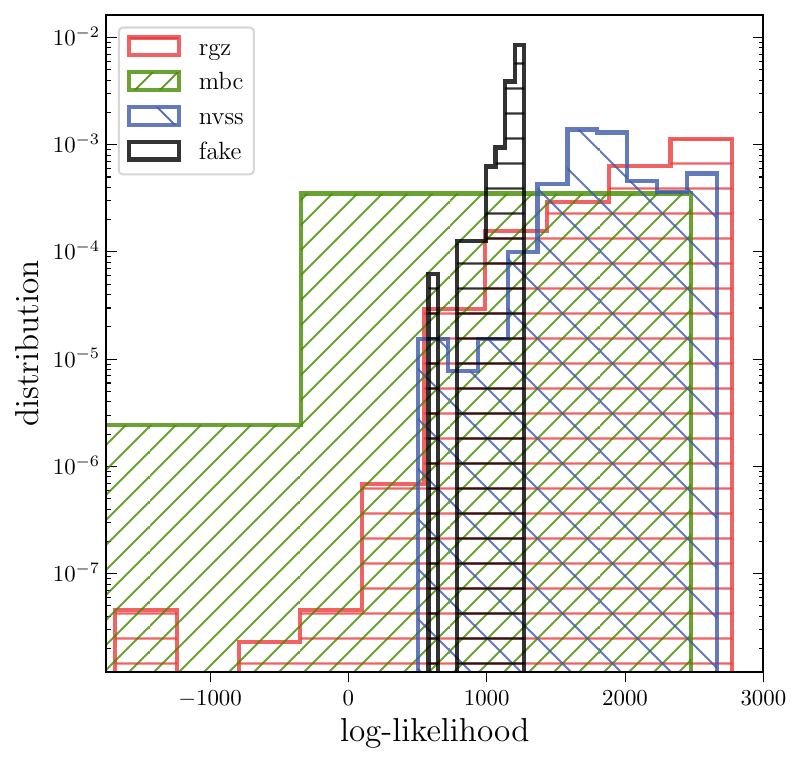}}
\cprotect\caption{Histograms of log-likelihood of all samples in each dataset, RGZ DR1 (red), MBC (green) and FR-DEEP NVSS (blue). The log-likelihood distribution of new images generated by \verb|VDVAE| is shown in solid black line.}
\label{loglikelihood}
\end{center}
\vskip -0.2in
\end{figure}
\begin{figure}[ht]
\vskip 0.2in
\begin{center}
\centerline{\includegraphics[width=1\columnwidth]{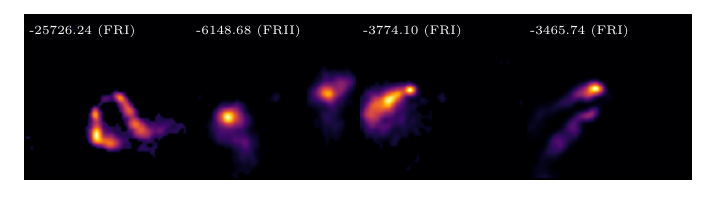}}
\caption{Out-of-distribution examples from MBC dataset based on the value of their loglikelihood which is the number presented on top of each image. The class of each galaxy from the MBC data is provided within round brackets.}
\label{ood-samples}
\end{center}
\vskip -0.2in
\end{figure}
\begin{figure}[ht]
\vskip 0.2in
\begin{center}
\centerline{\includegraphics[width=1\columnwidth]{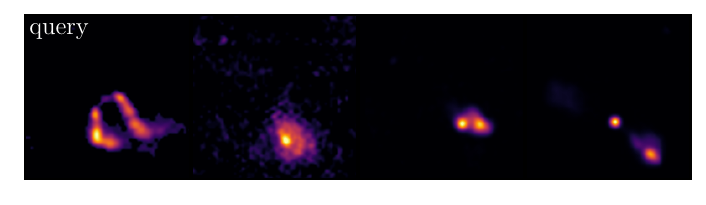}}
\caption{Search for images that are semantically similar to the an out-of-distribution sample (from MBC) in the RGZ DR1 data.}
\label{ood-search}
\end{center}
\vskip -0.2in
\end{figure}
This formulation allows the density estimation of the data after training. To generate a new data point $\textbf{x}_{\rm new}$, the method samples a point $\textbf{u}$ from the Gaussian prior and uses the mapping $f$. The density $p(\textbf{x})$ can be expressed as a product of conditionals $p(\textbf{x}) = \prod_{i} p(x_{i}|\textbf{x}_{1:i-1})$, parameterized as Gaussians, such that the $i$th conditional is given by \cite{papamakarios2017masked}
\begin{equation}
    p(x_{i}|\textbf{x}_{1:i-1}) = \mathcal{N}(x_{i}| u_{i},({\rm exp}\>\alpha_{i})^{2}),
\end{equation}
where $u_{i}$ and $\alpha_{i}$ are computed using scalar functions, $u_{i} = f_{u_{i}}(\textbf{x}_{1:i-1})$ and $\alpha_{i} = f_{\alpha_{i}}(\textbf{x}_{1:i-1})$. The scalar functions ($f_{u_{i}}, f_{\alpha_{i}}$) are constructed using Masked Autoencoder for Distribution Estimation (\verb|MADE|) \cite{germain2015made} which consists of dense layers. The autoregressive property is fulfilled by using appropriate masking, and making a conditional at $i$th \verb|MADE| layer dependent on the previous one $i$-\textit{1}th. In other words, \verb|MAF| architecture is built by chaining up several \verb|MADE| layers.
There are several flow based models depending on how the invertible function is constructed, such as \verb|Real NVP| \cite{dinh2016density}, but in our study, we train \verb|MAF| on the latent codes\footnote{Representations learned by \protect\Verb+VDVAE+.}.
\subsection{Log-likelihood of the data}
We consider a \verb|MAF| which comprises 48 \verb|MADE| blocks, each block composed of 2 fully connected layers of 512 hidden neurons. We choose Adam optimizer with learning rate of 0.0005 and train the \verb|MAF| model for 600 epochs on the latent codes of RGZ DR1 data. After training, we compute the log-likelihood of the representations of RGZ DR1, MBC and FR-DEEP NVSS and those of the new images generated by the decoder. Figure \ref{loglikelihood} shows the the log-likelihood histogram of each example in each dataset. The red, green, blue and black denote the log-likelihood distributions of  RGZ DR1, MBC, FR-DEEP NVSS and fake images respectively. The fact that the support of the FR-DEEP NVSS log-likelihood distribution is a subset of the RGZ DR1 base distribution indicates that FR-DEEP NVSS instances are not out-of-distribution (\verb|OOD|) with respect to RGZ DR1 dataset. In other words, the results suggest that the examples in both datasets are drawn from the same underlying distribution\footnote{Here we refer to the actual data distribution, not the log-likelihood distribution.}. However, it appears that some instances from the MBC data are considered \verb|OOD| with respect to RGZ DR1, as demonstrated by some log-likelihood scores that are outside the support of the RGZ DR1 log-likelihood distribution. We present in Figure \ref{ood-samples} instances that are associated with the lowest log-likelihood scores which, along with the class\footnote{This is given by the label of the MBC dataset.} (withing round brackets), are provided on the top left corner of each panel. For example, the image shown on the first panel from the left, which has the lowest log-likelihood, appears to be a bent-tail galaxy whose jets are bent. 
The second panel, labeled as FRII, shows two bright lobes which do not appear to be from the same central galaxy based on the diffuse structure surrounding each of them. The third and fourth panels present a core with one-sided jet and a bright spot seemingly disconnected from a nearby faint object respectively. 
Provided that the learned representations can be utilized to retrieve similar images in a dataset, we search for images in RGZ DR1 that are semantically similar to the outlier\footnote{Which is an instance in MBC.} corresponding to the lowest log-likelihood (top left panel in Figure \ref{ood-samples}). Search results are shown in Figure \ref{ood-search}. On the one hand it is clear that none of the retrieved images are semantically similar to the query, demonstrating the efficiency of the density estimator to assign low log-likelihood to images with features that haven't been seen during its training\footnote{Here we refer to the training of the density estimator.}. On the other hand, interestingly, it can be noticed that, like the query, each galaxy image in Figure \ref{ood-search} is located on the bottom right corner of the panel. This shows that, although the patterns are not similar, the feature components of the latent codes are such that the group of pixels that carries most of the information is roughly located at the same corner in each panel of Figure \ref{ood-search}. This test further demonstrates the meaningfulness of the learned representations. Lastly, Figure \ref{loglikelihood} also implies that the decoder is able to mimic RGZ DR1 data, as evidenced by the log-likelihood of each generated image that is well within the log-likelihood value range of the RGZ DR1 dataset.


\section{Conclusion}\label{sec:conclusion}
We have shown in this work that it is possible to learn meaningful latent codes of radio galaxy images that can be leveraged for some downstream tasks. We have trained on an unlabeled dataset a variant of Variational AutoEncoder (\verb|VAE|), whose both approximate posterior and prior are more expressive (compared to a vanilla \verb|VAE|) by resorting to a hierarchical structure composed of many stochastic layers of latent variables. We have assessed the overall performance of our \verb|VAE| model by looking at its ability to reconstruct the inputs, and analyzing how meaningful the representations it has learned during training are. In our investigation, we have also trained various \verb|SSL| based methods, \verb|SimCLR|, \verb|BYOL|, \verb|SimSiam|, and compared their performance in terms of classifying galaxies from labeled datasets with that of our model.
The features extracted by each model are visualized in a two dimensional subspace by using t-SNE, a dimensionality reduction method. To investigate if the learned representations from different models carry meaningful information,
six different classifiers -- $k$-nearest neighbors (\verb|knn|), random forest (\verb|rf|), support vector
machine (\verb|svc|), logistic regression (\verb|lr|), gradient boosting (\verb|grad|), and extra trees (\verb|ext|) -- are trained on them in order to identify FRI/FRII galaxies from two different datasets. Similarity search, which is another downstream task employing the compressed data, has also been conducted. Although the capacity of the \verb|VDVAE| model to generate new samples is not our primary objective in this work, we have checked how good it emulates the training data. Furthermore, we have estimated the log-likelihood of data by training a Masked Autoregressive Flow (\verb|MAF|), a state of the art density estimator, on the latent codes. This is especially useful in the context of finding anomaly/novelty in a dataset. We summarize our findings as follows:
\begin{itemize}
    \item Results suggest that our model is able to recover the inputs, capturing features like jet and diffuse structure, which indicates that the reconstructed images don't seem to suffer from blurriness, a known issue with \verb|VAE| models in general.
    \item The galaxy representations obtained from each model are well clustered with respect to angular size, implying that each method has properly learned to encode the high dimensional data.
    \item In a setup, the representations of galaxies from a labeled dataset, either MBC or FR-DEEP NVSS, are retrieved by a feature extractor (\verb|VDVAE|, \verb|SimCLR|, \verb|BYOL| or \verb|SimSiam|) and used to train several non-neural network  classifiers. In general, for MBC dataset, the information carried by the features extracted by the generative model has slightly better quality compared to those by the \verb|SSL| based models. The four best classifiers trained on the \verb|VDVAE| latent codes -- all achieving \textit{accuracy} $\ge 76\%$, \textit{roc-auc} $\ge 0.86$, \textit{specificity} $\ge 0.73$ and \textit{recall} $\ge 0.78$ -- outperform all the best classifiers of other setups in this work. The results on classifying galaxies in MBC dataset using learned representations also show that the performance of our generative model is comparable to that of the model in \cite{slijepcevic2022learning}. The top classifiers in all setups perform equally well on the FR-DEEP NVSS dataset. Interestingly, the performance of simple classifiers in our analyses is on par, if not better, with that of a CNN based model used in \cite{tang2019transfer}. This shows how meaningful the learned representation is.  
    \item The learned representations can be used for similarity search, as evidenced by the retrieved images that are semantically similar to the query image. We carry out searches for galaxies with large and small angular sizes. The results in both cases are consistent in the sense that all the images found exhibit similar patterns. In addition, the inclination of the galaxy in the query image is roughly found in all galaxies returned by the search. The importance of this application was highlighted in \cite{stein2021self}, where the encoded features were leveraged to search for similar images in a large dataset.
    \item We find that the decoder is capable of generating new images that are comparable with the training data overall. Neverthless, the generated images tend to be of smaller angular size, which can attributed to the bias in the dataset. The possibility that the decoder still lacks power, and hence requires more fine-tuning, can not be ruled out. However, a sufficiently powerful decoder is prone to a posterior collapse \cite{alemi2018fixing, chen2016variational} where the latent codes is no longer useful as they haven't been learned by the model. Provided that the main objective in this study is to learn the latent codes, increasing the power of the decoder in order to optimize the ability of the model to generate fake images needs to be approached carefully.
    \item The galaxies in FR-DEEP NVSS appear to have been drawn from the same distribution as those in RGZ DR1 dataset. This is evidenced by the log-likelihood values of the former which lie within the range of those from the latter. However, some galaxies withing MBC dataset are associated with log-likelihood scores outside RGZ DR1 base distribution\footnote{Here we refer to the distribution of log-likelihood.}, and therefore are considered \verb|OOD| (solely based on the likelihood as a metric). As a way to further validate both the usefulness of the latent codes and the density estimation by the \verb|MAF| model, we search for images in RGZ DR1 that are semantically similar to the \verb|OOD| instance (from MBC) associated with the lowest log-likelihood. We find that the search fails to return similar galaxies, corroborating the fact that the query image is indeed an \verb|OOD| with respect to RGZ DR1 dataset. It is also found that the new images generated by the decoder are \verb|in-distribution| with respect to RGZ DR1, as the estimated log-likelihood of each new instance is well within the log-likelihood distribution of RGZ DR1. It is worth noting that although both the \verb|VDVAE| encoder and decoder are trained simultaneously on the RGZ DR1 data, the new images which are obtained by sampling from latent space and reconstruction via the decoder have never been seen by the encoder which extracts the latent codes. This shows that the decoder is able to emulate the RGZ DR1 data. 
\end{itemize}
The generative model in this work has shown promising performance. For future investigation one question that can be addressed is the impact of the input dimensions on the results, for instance by considering $128\times 128$ pixels resolution of the images used for training. This is one way to assess the robustness of the method.  

\acknowledgments
SA acknowledges financial support from the {\it South African Radio Astronomy Observatory} (SARAO). SA is grateful to both Anna Scaife and Inigo Val Slijepcevic for the very helpful discussions about the data. HT gratefully acknowledges support from the Shuimu Tsinghua Scholar Programme of Tsinghua University, the China Postdoctoral Science Foundation fellowship 2022M721875, and long-lasting support from various machine learning groups notably the University of Manchester Jodrell Bank Centre for Astrophysics, the TAGLAB research group, and the DoA at Tsinghua.

This publication has been made possible by the participation of more than 250,000
volunteers in the Galaxy Zoo Project. The data in this paper are the result of the efforts of
the Radio Galaxy Zoo volunteers, without whom none of this work would be possible.
Their efforts are individually acknowledged at \href{http://rgzauthors.galaxyzoo.org}{http://rgzauthors.galaxyzoo.org}.


\bibliographystyle{JHEP}
\bibliography{rgz}

\end{document}